\documentstyle[12pt]{article}
\begin{document}
\begin{center}
{\Large Torsion Modified Plasma Screening in Astrophysics}
\vspace{1cm}

\noindent
C. Sivaram\footnote{Indian Institute of Astrophysics, Bangalore India, and
Universita degli Studi di Ferrara, Dipartimento di Fisica, Italia.} and L.C.
Garcia de Andrade\footnote{Instituto de F\'{\i}sica , UERJ, Rua S\~{a}o
Francisco Xavier 524, Rio de Janeiro Brasil.}
\end{center}
\vspace{2cm}
\begin{center}
{\Large Abstract}
\end{center}
\vspace{0.5cm}

The torsion modified Maxwell-Proca equations when applied to describe
a plasma is shown to lead to a correction to the Debye screening
length.

For hot new born neutron stars the torsion correction is shown to be
significant. This effect may provide an indirect evidence for Torsion.
\newpage

In recent papers\cite{1,2}, the Maxwell equations in the back ground
of an Einstein-Cartan spacetime were shown to take the form
\begin{eqnarray}
\nabla . \vec{E} &=& 4 \pi \rho - \frac{3\lambda}{k}
\left({\partial}_{0} Q^{0} + \nabla . \vec{Q} \right) \phi \label{1} \\ 
\frac{\partial \rho}{\partial t} &+& \nabla . \vec{J} = - \frac{3
\lambda}{4 \pi k} \left( \frac{{\partial}^{2} Q^{0}}{\partial
t^{2}}\right) \phi \label{2} \\
\nonumber
\end{eqnarray}
Here Q is the torsion vector, which is the background spin density of
$ \displaystyle \sigma $ given by the Einstein-Cartan theory as
\begin{equation}
Q = \frac{\displaystyle 4 \pi G \sigma}{\displaystyle c^{\displaystyle 2}}
\label{3}
\end{equation}
In the gauge $\displaystyle \nabla . \vec{Q} =0$, the first of above
equations is equiva lent to the Proca-Equation, ie,
\begin{equation}
\nabla . \vec{E} = 4 \pi \rho - {m_{\displaystyle \gamma}}^{2}\ \phi
\label{4}
\end{equation}
where $\displaystyle {m_{\gamma}}^{2}$ is given in this case by:
\begin{equation}
{m_{\gamma}}^{2} = \frac{3 \lambda}{k} \left( \frac{{\partial
Q_{0}}}{\partial t}\right)
\label{5}
\end{equation}
In the Work\cite{3}, which involved the torsion-photon coupling of
the form $\displaystyle \cong \lambda R(\rho ) A_{\mu}\ A^{\mu}$, it
was shown
\begin{equation}
{m_{\gamma}}^{2} \cong \lambda \ Q^{2}
\label{6}
\end{equation}
being constraint by\cite{4} $\displaystyle \lambda < 10^{-24}$.

\noindent
We now apply the above equations to Debye sreening in a plasma. We
write eq.(\ref{1}) in the form:
\begin{equation}
{\nabla}^{2} \phi = 4 \pi n_{e} e - {m_{\gamma}}^{2} \phi
\label{7}
\end{equation}
Where $\displaystyle n_{e}$ is the  electron number density,
$\displaystyle e $ is the electron charge in a hot charged plasma at
temperature T, is given by the Boltzmann distribution :
\begin{equation}
n_{\displaystyle e} \cong n_{\displaystyle e_{0}} exp {\left(- e \phi / k_{B}
\ T \right)}
\label{8}
\end{equation}
Where $\displaystyle k_{\displaystyle B}$ is the Boltzmann constant,
$\displaystyle \phi $ is the average potential in the vicinity of an ion, so
that the typical electron has an electros static energy
$\displaystyle \cong e \phi $
\newpage

As is usual in the discussions of plasma screening we can approximate
from eq. (\ref{8}) as
\begin{equation}
\displaystyle n_{\displaystyle e} \cong \displaystyle
n_{\displaystyle e_{0}} \left( 1 - \displaystyle e \phi / k_{B}T \right)
\label{9}
\end{equation}
Eq. (\ref{7}) then has the form
\begin{eqnarray}
{\nabla}^{2} \phi &=& 4 \pi \displaystyle n_{\displaystyle
e_{0}}\displaystyle e\ exp \left( - \displaystyle e \phi /k_{B} T\right) - {m_{
\displaystyle \gamma}}^{2} \phi \label{10} \\
{\nabla}^{2} \phi &\cong & 4 \pi \displaystyle n_{\displaystyle
e_{0}}\displaystyle  exp \left(1 - \displaystyle e \phi /k_{B} T\right) - {m_{
\displaystyle \gamma}}^{2} \phi \label{11} \\
\nonumber
\end{eqnarray}
with $\displaystyle {m_{\gamma}}^{2} \cong \lambda Q^{2}$ (in the
static case considered here )

\noindent
As shown in refs. \cite{4,5}, the contraint $\displaystyle \lambda $
is $\displaystyle \cong 10^{-24}$.

\noindent
The solution of eq.(\ref{1}) can in general be written as 
\begin{equation}
\phi \cong \frac{\displaystyle e}{\displaystyle r} \left( 1 - exp \left(
- {\lambda}^{' - 1}_{D \ \displaystyle \gamma}\right) \right)
\label{12}
\end{equation}
Where the modified Debye lenght is given as
\begin{equation}
{\lambda}^{'}_{D} \cong \left(\frac{k_{B} T}{4 \pi n_{e_{0}} e^{2}} +
\frac{{\hbar}^{2} }{\lambda Q^{2} c^{2}}\right)^{1/2}
\label{13}
\end{equation}
or for a varying torsion background
\begin{equation}
{\lambda}^{'}_{D} \cong \left(\frac{k_{B} T}{4 \pi n_{e_{0}} e^{2}} +
\frac{{\hbar}^{2}k}{3 \lambda \left(\frac{\displaystyle \partial Q_{0}}{
\displaystyle \partial t}\right)c^{2}}\right)^{1/2}
\label{14}
\end{equation}
In the limit of zero torsion, eqs. (\ref{13}) and {(\ref{14}) give the
usual Debye screening in plasmas.

Since the effects of the torsion scale with the spin density. We
would expect this to be significant in neutron stars with a high
number density such as $\displaystyle n_{0} \cong 10^{40}{cm}^{-3}$.

We can estimate the temperature at which  the torsion screening term
becomes comparable to the usual Debye term.

Since $\displaystyle Q \cong \frac{\displaystyle 4 \pi G \ n_{0} \hbar}{
\displaystyle c^{2}}$ we get the corresponding temperature $T_{0}$ at
which the two terms are comparable as 
\newpage
\begin{equation}
T_{0} \cong \frac{c^{4} \ e^{2} \beta}{4\pi G^{2}\ \lambda \ n_{0} k_{B}}
\label{15}
\end{equation}
For $\displaystyle n_{0} \cong 10^{40} {cm}^{-3}$ this gives $T_{0}
\cong 10^{12}k$.

This shows that for sufficiently hot new born neutron stars\cite{5}
(X ray emiting neutron star sources typically have $T_{0}$ of a few
million degrees), the torsion modified electrodynamic terms. Here we
assume $\displaystyle \lambda \cong 1$ i.e. strong torsion neutron spin
coupling. Instead of eq.(\ref{3}) we could also use for Q the expression
\begin{equation}
Q = - \nabla \ln (1 + \lambda A^{2} )
\label{16}
\end{equation}
This would modify the equation (\ref{4}) as :
\begin{displaymath}
{\nabla}^{2}\phi = 4\pi \rho - (\nabla \phi )^{2} \lambda \rightarrow
\phi \cong \frac{A}{r} + B \ln (a - r)
\end{displaymath}
(A and B constants). We would still get the screening solution but with an 
extra logarithimic term which does not contribute to the screening. The
consequences of this for the stability of the system would be investigated in 
another work\cite{6}. We also like to mention that this is not the
first time massive photons have been assumed to "surround" a neutron
star. In 1969 Feinberg\cite{7} have assumed that all dispersion in
the Crab pulsar results from massive photons which allowed him to put
a limit of $m_{\gamma} < 10^{- 44}g$ on the photon mass. Finally it
is interesting to note that although direct effects F. Torsion in
neutron star like the contribution to eccentricity\cite{8} are too
small to be measurable, indirect effects of Torsion in neutron star
like the one proposed here and cumulative effects like the ones
proposed by Zhang\cite{9,10} and his group can be useful as an
indirect evidence for Torsion Gravity.

\end{document}